\newif\ifarXiv
\arXivtrue
\documentclass[runningheads]{llncs}
\usepackage[T1]{fontenc}
\usepackage{graphicx}

\makeatletter
\renewcommand\subsubsection{%
    \@startsection{subsubsection}{3}{\z@}%
        {-5\p@ \@plus -1\p@ \@minus -1\p@}%
        {-0.5em \@plus -0.22em \@minus -0.1em}%
        {\normalfont\normalsize\bfseries\boldmath}}
\makeatother
\newcommand\ITEM[1]{\par\smallskip\noindent\textit{#1}}

\usepackage{cite}
\usepackage{enumitem}
\usepackage{amsmath,amssymb,amsfonts}
\usepackage{algorithmic}
\usepackage{textcomp}
\usepackage{xcolor,colortbl}
\definecolor{grey1}{gray}{0.9}
\definecolor{grey2}{gray}{0.7}
\usepackage{fancyvrb}
\usepackage{array}
\newcolumntype{L}[1]{>{\raggedright\let\newline\\\arraybackslash\hspace{0pt}}m{#1}}
\newcolumntype{B}[1]{>{\raggedright\let\newline\\\arraybackslash\hspace{0pt}}b{#1}}
\newcolumntype{C}[1]{>{\centering\let\newline\\\arraybackslash\hspace{0pt}}m{#1}}
\newcolumntype{R}[1]{>{\raggedleft\let\newline\\\arraybackslash\hspace{0pt}}m{#1}}
\usepackage{pgfplots}
\usepackage{pgf}
\usepackage{pgfplotstable}
\usepackage{tikz}
\usetikzlibrary{positioning,calc}
\usepackage{rotating}
\usepackage{multirow}
\usepackage{booktabs}
\usepackage[font=small,labelfont=bf,tableposition=top]{caption}
\usepackage{xfrac}
\usepackage{longtable}
\newcommand\SW[1]{\begin{turn}{90}#1\end{turn}}
\usepackage{bbding}

\newcommand\YES{\Checkmark}
\newcommand\med{$\circ$}
\newcommand\good{$+$}
\newcommand\bad{$-$}
\newcommand\WA{\cellcolor{pink}} 

\newcommand\op[1]{{\textsc{#1}}}
\usepackage[breaklinks=true]{hyperref}
\usepackage{breakurl}

\begin{document}

\title{Consolidation of Ground Truth Sets \\ for Weakness Detection in Smart Contracts%
\ifarXiv\thanks{Extended version of a paper presented at the Workshop on Trusted Smart Contracts 2023, to be published in LNCS, Springer.}\fi}
\titlerunning{Consolidated Ground Truth}
\author{Monika di Angelo\orcidID{0000-0002-4217-4530} \and
Gernot Salzer\orcidID{0000-0002-8950-1551}}
\authorrunning{M. di Angelo, G. Salzer}
\institute{TU Wien, Vienna, Austria\\
\email{\{monika.di.angelo, gernot.salzer\}@tuwien.ac.at}
}
\maketitle

\begin{abstract}
  Smart contracts are small programs on the blockchain that often handle valuable assets. Vulnerabilities in smart contracts can be costly, as time has shown over and over again.
  Countermeasures are high in demand and include best practice recommendations as well as tools supporting development, program verification, and post-deployment analysis.
  Many tools focus on detecting the absence or presence of a subset of the known vulnerabilities, delivering results of varying quality.
  Most comparative tool evaluations resort to selecting a handful of tools and testing them against each other.
  In the best case, the evaluation is based on a smallish ground truth.
  For Ethereum, there are commendable efforts by several author groups to manually classify contracts.
  However, a comprehensive ground truth is still lacking.

  In this work, we construct a ground truth based on publicly available benchmark sets for Ethereum smart contracts with manually checked ground truth data.
  We develop a method to unify these sets.
  Additionally, we devise strategies for matching entries that pertain to the same contract, such that we can determine overlaps and disagreements between the sets and consolidate the disagreements.
  Finally, we assess the quality of the included ground truth sets.
  Our work reduces inconsistencies, redundancies, and incompleteness while increasing the number of data points and heterogeneity.

\end{abstract}

\section{Introduction}
To support the development of Ethereum smart contracts (SCs) and to analyze SCs that have been deployed, over 140 tools were released until mid-2021~\cite{Rameder2022}, and new tools keep appearing. 
The sheer number of tools makes it difficult to choose an appropriate one for a particular use case.
Moreover, it is difficult to assess the effectiveness of the many methods proposed, and to judge the relevance of various extensions.
Tool comparisons can facilitate the selection process.
However, many tool surveys are based on academic publications that focus on the methods employed by the tools, or on whitepapers of the tools themselves.
For a thorough quality assessment of the tools, it is necessary to also install and systematically test the tools -- preferably with an appropriate ground truth set of SCs.

Given the scarce availability of an appropriate ground truth, tool developers adopted the practice of comparing their tool to previous ones, often with the somewhat biased intention of demonstrating the superiority of their tool in a particular respect.
This approach is justified by the need for an evaluation despite the lack of an established ground truth.
However, there are major concerns about the validity of such evaluations.

\ITEM{Undetermined quality of tools:} Since the point of reference is unclear, a comparison to something of unknown quality only provides relative information. 

\ITEM{Dependence between tools:} When a new tool builds on tools published earlier, there is a tendency to compare it to exactly those tools in order to show the improvements.
With the quality of the base tool(s) not clearly determined, the relative quality assessment remains vague.

\subsubsection{Ground truth (GT).}
In our context, a \emph{ground truth} for a particular program property is a set of smart contracts (given as source or bytecode) together with assessments that state for each contract whether it satisfies the property or not. 
As the term truth suggests, these assessments are supposed to be definitive and reliable. 
To foster trust into the ground truth, it may be accompanied by a specification of the process how the assessments were obtained (e.g.\ by expert evaluation) or by objective arguments for the assessments (e.g.\ by specifying program inputs that solicit behavior satisfying the property, or by showing that such inputs do not exist).

\subsubsection{Goals and approach.}
The primary goal of this work is to compile a unified and consolidated ground truth of SCs with manually labeled properties, starting from GT sets that are publicly available and documented.
Ultimately, we aim at a uniformly structured collection of contracts with verified properties that harnesses the individual efforts that have been invested into the original datasets.

\ITEM{Unification.}
From related work, we collect benchmarks containing GT data.
We extract information on the corresponding contracts (like address, source code, bytecode, location of the issue) as well as classifications (properties tested, assessments) and introduce a unique reference for every entry in the original dataset.
We clean the data by repairing obvious mishaps and complete it using our database of source codes and chain data.

\ITEM{Consolidation.}
To consolidate the datasets, we introduce four attributes per contract: the address (with chain and creation block) if the contract has been deployed, as well as unique fingerprints of the source code, the deployment and the deployed bytecode.
Based on these attributes, we determine and eliminate discrepancies within the individual datasets.
Then, we map the classifications to a common frame of reference, the SWC\footnote{\url{https://swcregistry.io}} classes and DASP\footnote{\url{https://dasp.co/}} scheme.
Relying again on the attributes, we determine overlaps between datasets, detect disagreements, and examine their cause.

\ITEM{Quality assessment.} Based on the taxonomy by Bosu et al.~\cite{Bosu2013taxonomy} for assessing data quality in software engineering, we assess the included GT sets set with regard to the three aspects accuracy, relevance, and provenance. 
For accuracy, we consider incompleteness, redundancy, and inconsistency; for relevance, we consider heterogeneity, amount of data, and timeliness; for provenance, we consider accessibility and trustworthiness.

\section{Definition of Terms}

To discuss the data, we use the following terms. 

\ITEM{Property, weakness, vulnerability:}
Most contract properties addressed in datasets constitute program weaknesses, with a few exceptions like honeypots.
In software engineering at large, vulnerabilities are weaknesses that can be actually exploited, while blockchain literature tends to use the two terms synonymously.
Throughout the paper, we prefer the term \emph{weakness}, and use \emph{property} for general statements. 

\ITEM{Judgment:} If a property holds, the corresponding judgment is ``positive''. If a property does not hold, the judgment is ``negative''. If the assessment is inconclusive or does not make sense, the judgment is ``not available'' (n/a).

\ITEM{Assessment:} a triple consisting of a contract, a single property, and a judgment of the latter in the context of the former.

\ITEM{Entry:} smallest unit of a dataset according to its authors. Depending on the organization of the dataset, an entry consists of a single assessment or of multiple assessments pertaining to the same contract. We use the term mainly to relate to the original publications accompanying the datasets.

\ITEM{Contradiction:} a group of two or more assessments for the same contract and property, but with conflicting judgments.

\ITEM{Duplicates:} multiple assessments for the same contract and property with identical judgments.

\section{Benchmark Sets with Ground Truth Data}\label{sec:benchmarks}
In this section, we specify the selection of the benchmarks sets and give an overview of the contents in the included sets.

\subsection{Selection of GT Sets}
From the systematic literature review~\cite{Rameder2022}, where Rameder et al.\ identified benchmark sets of smart contracts for the quality assessment of approaches to weakness or vulnerability detection, we extracted all references that contain a ground truth.
Moreover, for the years 2021 and 2022, we searched for further GT sets.

\ITEM{Inclusion criteria.}
We include all sets that provide a ground truth by either manually checking the contracts or by generating them via deliberate and systematic bug injection.

\ITEM{Exclusion criteria.}
We omit sets that reuse the samples of other sets without contributing assessments of their own.
Moreover, we exclude sets having been assessed automatically, e.g.\ by combining the results of selected vulnerability detection tools by majority voting.
While they may constitute interesting test data, they do not qualify as a ground truth.
See section~\ref{sec:excluded} in the appendix for excluded sets.

\ITEM{Overview of included sets.}
Table~\ref{tab:gt_included} lists the sets finally selected as the basis of our work.
\emph{Wild} sets contain contracts that have been actually deployed on the main or a test chain. \emph{Crafted} sets have been engineered to exemplify typical weaknesses or generated by injecting bugs into source code.

\subsection{Structure of the Included Sets}

The datasets differ regarding the number of assessments per entry, the identification of contracts, the way assessments are specified, and the information provided per contract.
Table~\ref{tab:gt_contents} in the appendix gives an overview.

\ITEM{Identification:}
Usually, contracts are identified either by a file with the Solidity source or by a chain address.
Only one dataset specifies just an internal identifier, which in most cases contains an address.

\ITEM{Assessments:}
The majority of datasets provides the assessments in a structured form as \texttt{csv}, \texttt{json}, \texttt{xlsx} or \texttt{ods} files.
Five datasets encode the weakness and partly also the judgment in the filepath or use prose.

\ITEM{Contract information:}
The datasets may provide chain address, Solidity sources from Etherscan or elsewhere, deployment and/or runtime bytecode.

\subsubsection{Crafted and wild sets.}
Depending on the provenance of the contracts, we divide the datasets into two groups.
The \emph{wild} group comprises eight collections of contracts that have been deployed either on the main or a test chain, hence they all provide chain addresses or source code from Etherscan.
The \emph{crafted} sets contain at least some contracts that have not been deployed to a public chain\footnote{%
  The distinction between \emph{crafted} and \emph{wild} sets is not strict.
  Crafted sets may contain some contracts from public chains in modified
  or unmodified form.}.
One set has been obtained from the SWC registry, where it illustrates the SWC taxonomy.
Two sets, JiuZhou and SBcurated, are related to tool evaluations.
The set NotSoSmartContracts is intended for educational purposes, and the set SolidiFI was generated from Solidity sources by injecting seven types of bugs.

\subsection{Methods Employed in the Included Sets}

Table~\ref{tab:gt_methods} in the appendix briefly summarizes the methods that the authors claim to have applied for compiling their respective GT sets. 
The group of crafted sets are often collections of real mishaps, stripped versions to demonstrate bad practices, crafted to the point, or generated.

For the wild sets, the manual assessments were mainly done by inspecting the Solidity source code.
However, three sets contain a total of 277 entries with addresses for which no source code has been published.
The ground truth of eThor contains 118 bytecode-only contracts, of which a selection was inspected~\cite{Schneidewind2020}.
EverEvolvingGame lists 28 addresses without source code. However, the assessments are based on inspecting transactions rather than code~\cite{Zhou2020}.
For Zeus~\cite{Kalra2018}, Kalra et al.\ claim on page 12 that they have manually assessed all 1524 entries in the set for seven vulnerabilities each.

\subsection{Summary of Assessments in the Included Sets}

Table~\ref{tab:gt_included} gives an overview of the assessments in the sets.
The first column contains a reference to the publication presenting the sets, while the second one gives the number of entries.
The subsequent columns quantify the assessments, specifying the total number as well as a breakdown by judgment type.
The column for ignored assessments indicates the number of duplicate or contradicting assessments, as discussed in section~\ref{ssec:ignored}.

\begin{table}[!ht]
  \centering
  \setlength\tabcolsep{3pt}
  \caption{Included GT Sets.}\label{tab:gt_included}
  \newcolumntype{g}{>{\centering\arraybackslash\columncolor{grey1}}r}
  \begin{tabular}{lrr|rrrrgr|rr}
    \toprule
    \multicolumn{3}{c|}{~}&\multicolumn{6}{c|}{assessments} & \\
    Set
      & \SW{reference} & \SW{entries} &\SW{total} & \SW{positive} & \SW{negative} & \SW{n/a}
      & \SW{ignored} & \SW{unmapped} &\SW{weaknesses} &\SW{SWC classes} \\ 
    \midrule
    \href{https://github.com/Jiachi-Chen/TSE-ContractDefects}{CodeSmells}
      & \cite{Chen2020defining}        &  587 & 11740 & 2293 & 9073 & 374 & 1330 & 5870 & 20 & 10 \\
    \href{https://github.com/gongbell/ContractFuzzer/tree/master/examples}{ContractFuzzer}
      & \cite{Jiang2018ContractFuzzer} &  379 &   379 &  379 &    0 &   0 &    4 &    0 &  7 &  7 \\
    \href{https://drive.google.com/file/d/1k0Edw2r1Z59WBc8SFbeh85hJMydGNPGz/view}{Doublade}
      & \cite{Xue2019doublade}         &  319 &   319 &  152 &  167 &   0 &   40 &    0 &  5 &  5 \\
    \href{https://secpriv.wien/ethor/}{eThor}
      & \cite{Schneidewind2020}        &  720 &   720 &  196 &  512 &  12 &   18 &    0 &  1 &  1 \\
    \href{https://drive.google.com/file/d/1190VXwu502M-vgT8yyuFp0lFUVlxnMhO/view?usp=sharing}{EthRacer}
      & \cite{Kolluri2019}             &  127 &   127 &   69 &   47 &  11 &   16 &    0 &  2 &  1 \\
    \href{https://drive.google.com/open?id=1xLssDxYWyKFCwS5HUrQaSex0uwJRSvDi}{EverEvolvingG.}
      & \cite{Zhou2020}                &  344 &   344 &  344 &    0 &   0 &   52 &  271 &  5 &  3 \\
    \href{https://www.dropbox.com/sh/90tm5drmeep9bqy/AAB0jKxkIevNct2eIvsYb7Oqa?dl=0}{NPChecker}
      & \cite{Wang2019detecting}       &   50 &   250 &   28 &  222 &   0 &   31 &    0 &  5 &  5 \\
    \href{https://goo.gl/kFNHy3}{Zeus}
      & \cite{Kalra2018}               & 1524 & 10533 & 2726 & 7807 &   0 & 3210 &    0 &  7 &  7 \\
    \midrule
    \href{https://github.com/xf97/JiuZhou}{JiuZhou}
      & \cite{Zhang2020framework}      &  168 &   168 &   68 &  100 &   0 &    3 &   39 & 53 & 33  \\
    \href{https://github.com/crytic/not-so-smart-contracts/}{NotSoSmartC.}
      &                                &   31 &    34 &   24 &   10 &   0 &    0 &    2 & 18 & 12 \\
    \href{https://github.com/smartbugs/smartbugs-curated}{SBcurated}
      & \cite{Ferreira2020}            &  143 &   145 &  145 &    0 &   0 &   16 &    0 & 10 & 16 \\
    \href{https://github.com/DependableSystemsLab/SolidiFI-benchmark}{SolifiFI}
      & \cite{Ghaleb2020SolidiFI}      &  350 &   350 &  350 &    0 &   0 &    7 &    0 &  7 &  7 \\
    \href{swcregistry.io}{SWCregistry}
      &                                &  117 &   117 &   76 &   41 &   0 &    1 &    0 & 33 & 33 \\
    \bottomrule
\end{tabular}
\end{table}

To compare the weaknesses covered by the sets, we map the individual assessments to the taxonomy provided by the SWC registry (table~\ref{tab:coverage}).
Section~\ref{ssec:mapping} discusses the mapping in detail.
Properties not represented in the SWC registry remain unmapped, leading to unmapped assessments.
The last two columns of table~\ref{tab:gt_included} give the number of weaknesses as defined by the set and the number of covered SWC classes.
When the number of weaknesses is larger than the number of SWC classes covered, it either means that there are unmapped assessments or that several weaknesses are mapped to the same SWC class.
\section{Unified Ground Truth} \label{sec:unify}
In this section, we describe the process of merging the selected sets into a unified ground truth.
We extract relevant data items, assign unique identifiers to the entries, repair mishaps, normalize the data to obtain a common format, add missing information from other data sources and investigate data variability.

\subsection{Extracting Data from the Original Sets}

For each repository selected (section~\ref{sec:benchmarks}), we identify the parts pertaining to a ground truth, and use a Python script to extract relevant items.
At a minimum, we need information to identify a contract, a property, and a corresponding judgment.
Table~\ref{tab:gt_contents} in the appendix lists the different forms this information may take.

Most sets have not been designed for automated processing. They contain inconsistencies, errors, and information only intelligible to humans.
We encountered numerous invalid Ethereum addresses, inconsistent spellings, invalid data formats, and wrong information (like bytecode not corresponding to the given source code).
For the sake of transparency, we left the original sets unchanged and integrated the fixes into the Python scripts.

\subsection{Completing the Data} \label{ssec:completion}

To identify duplicate or contradicting assessments, and to arrive at a consolidated ground truth usable in different scenarios, each contract should be given by its source, contract creation, and runtime code as well as by its chain address (if deployed).
Most repositories contain only some of this information.
With the help of data from Ethereum's main chain and Etherscan's repository of source code, we were able to complete most missing data.

\ITEM{Contracts with addresses:}\,\footnote{%
Addresses by themselves are not sufficient to identify a contract.
Apart from information about the chain, we also need the deployment time if the contract or an ancestor is the result of a \op{create2} operation.
However, as the data in the repositories mostly predates the introduction of this operation, we encountered no contract of this type.
Hence, for our purposes knowing the address and chain is sufficient.
We only use the block numbers of deployments for analyzing changes over time.}
We query the respective chain for the bytecodes, and Etherscan for the source code (if available).

\ITEM{Contracts with source code:}
We use the fingerprint of the source code to look it up in an internal database.
If there is a match, we retrieve the deployment address and proceed as above.
Otherwise, the source code can be compiled to obtain the corresponding bytecode.
Given the variability of compilation, this step most likely will not result in code matching code obtained elsewhere, and is thus inferior when searching for duplicates.

\ITEM{Contracts with bytecode:}
The contracts considered here all come with an address or some source code.
However, to cross-check and to confirm guesses about the chain, we use fingerprints of any provided bytecode to look up public deployments.
Moreover, we extract the runtime code from given deployment code.

\subsubsection{Fingerprints.}
To detect identical contracts, we use fingerprints of the code.
For source code, we eliminate comments and white space before computing the MD5 hash.
A second type of fingerprint additionally eliminates \texttt{pragma solidity} statements prior to hashing.
For bytecodes, we replace metadata sections inserted by the Solidity compiler with zeros before computing the MD5 hash.

\subsection{Variability in the Unified GT Set} \label{ssec:variability}
We portrait the variability with regard to the contract language (Solidity or EVM bytecode) as well as the range and distribution of Solidity versions and time of deployment.

\begin{figure}
\centering
\includegraphics[width=.75\textwidth]{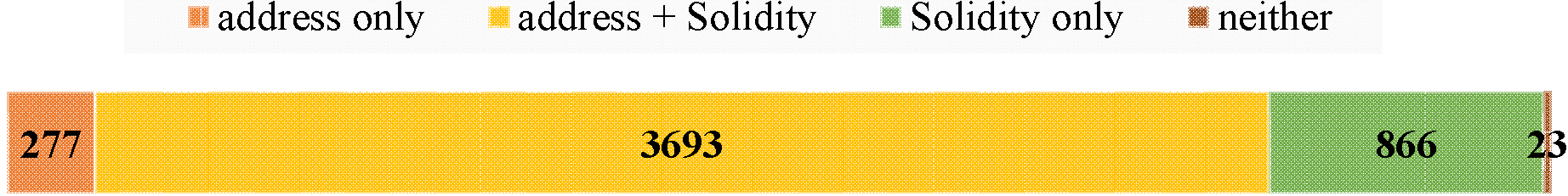}
\caption{%
Addresses (orange and yellow) and Solidity source files (yellow and green) in the entries in the unified GT set.}%
\label{fig:variability}
\end{figure}

\ITEM{Contract identification.}
We need some reference to a contract, be it an address or a source file.
Figure~\ref{fig:variability} depicts the number of entries in the unified GT set, for which we have an address, a source, both, or neither.

In the unified set, there are 4\,859 entries in total, of which 4\,559 (93.8\,\%) come with a Solidity source and 3970 (81.7\,\%) with a deployment address.
While 3693 (76.0\,\%) entries are associated with both, address and source, there are 866 (17.8\,\%) entries with a Solidity source only, and 277 (4.6\,\%), for which a source file is neither provided not can be retrieved.
This concerns 28 entries in the set EverEvolvingGame, 131 in Zeus, and 118 in eThor.
Moreover, 23 entries indicate neither an address nor a source file, but rather refer to a Solidity file without providing it (all in Zeus).

\ITEM{Chains.}
Entries with address refer to 2731 unique addresses, with the majority (2461) from the main chain, 268 from Ropsten, and one from Rinkeby.
For one address in Zeus, we were not able to locate it on any public chain.

\ITEM{Solidity versions.}
Solidity, the main programming language for smart contracts on Ethereum and beyond, has been evolving with several breaking changes so far.
In the included sets, we see predominantly versions 0.4.x as depicted in the left part of figure~\ref{fig:distribution_sol}.
While the versions 0.4.x were current throughout 2017 up to early 2018, versions 0.8.x started December 2020 and are still current in mid 2023.
The highest Solidity version in the GT sets is v0.6.4.

\bigskip
\noindent
\begin{minipage}{.48\textwidth}
\includegraphics[width=\linewidth]{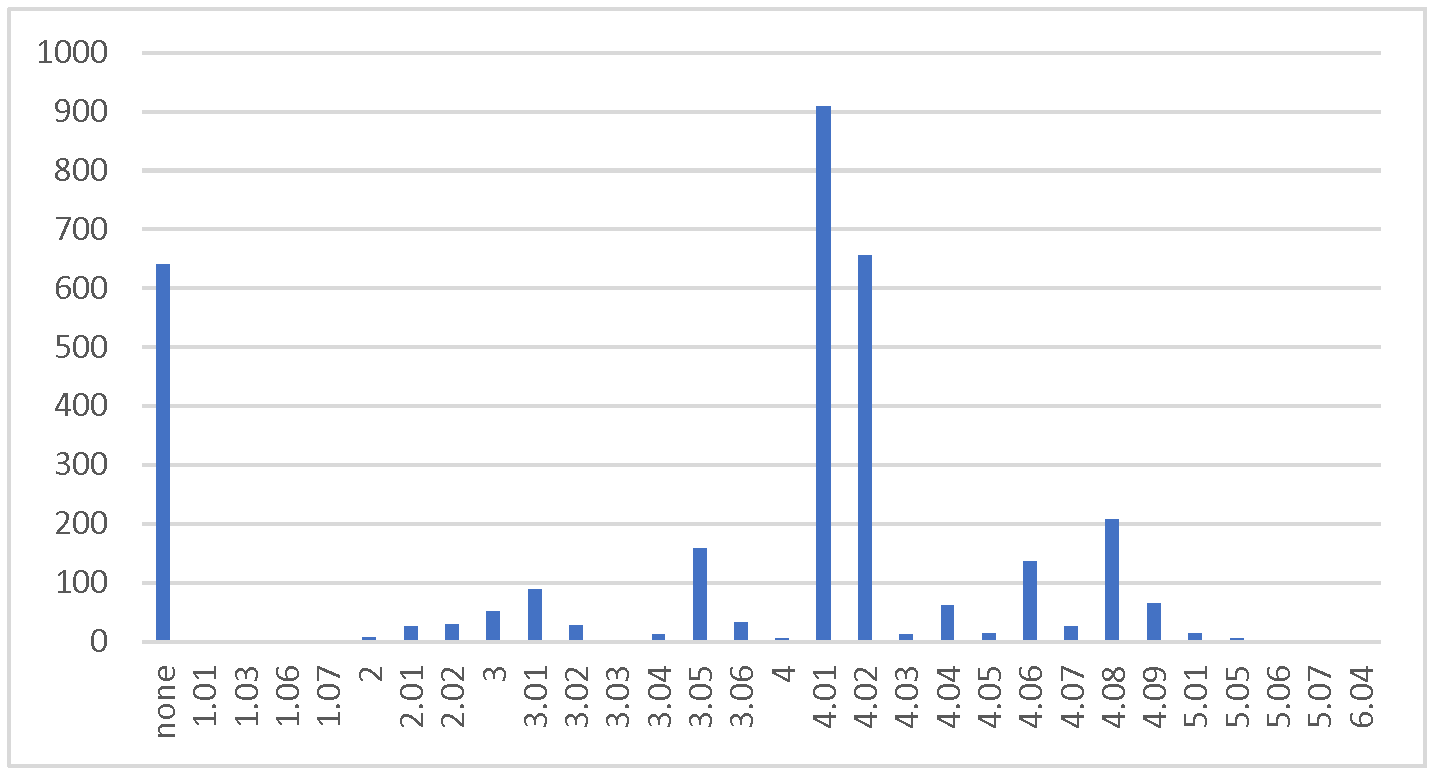}
\captionof{figure}{Distribution of Solidity versions in the included GT sets.\\~}
\label{fig:distribution_sol}
\end{minipage}\hfill%
\begin{minipage}{.48\textwidth}
\includegraphics[width=\linewidth]{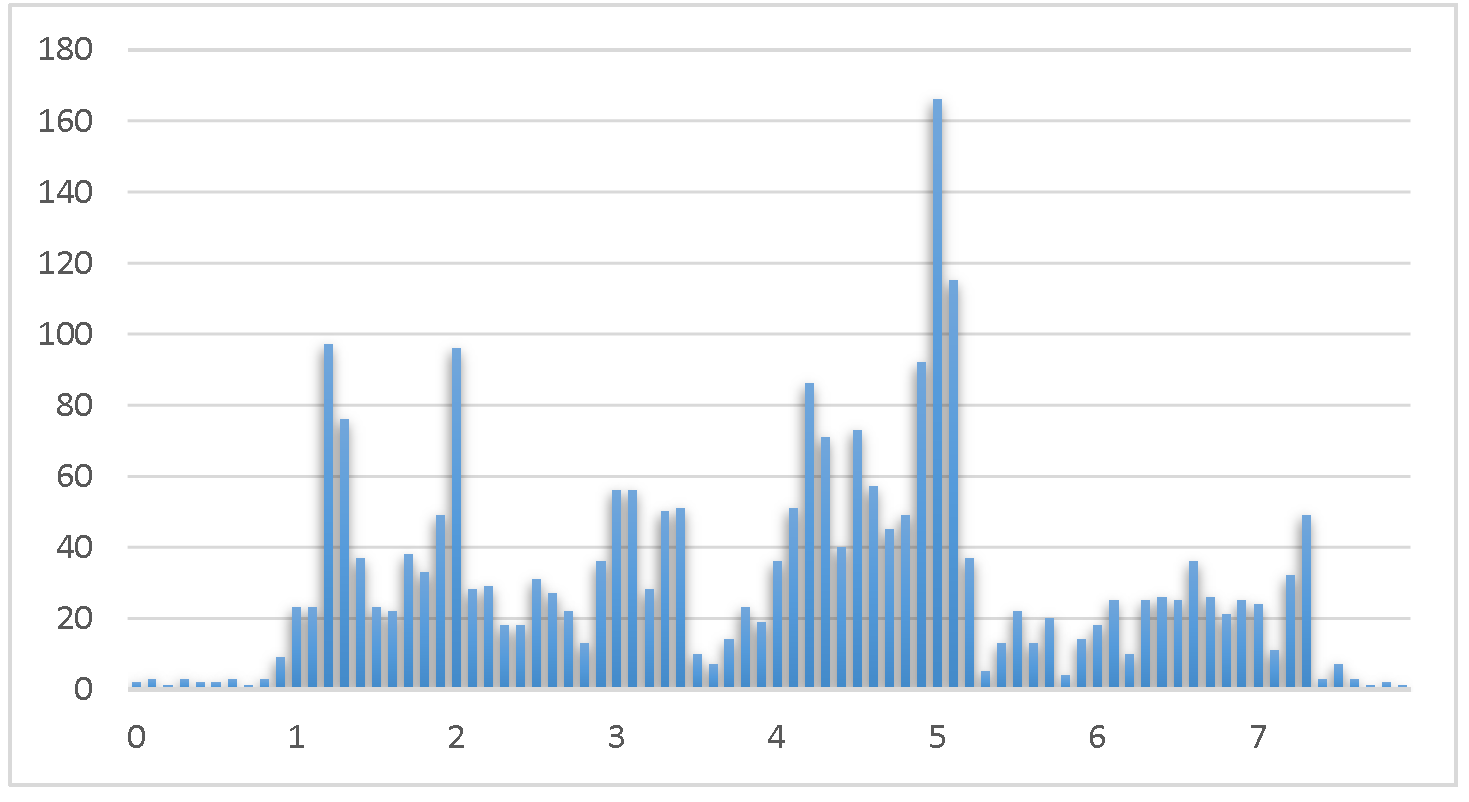}
\captionof{figure}{Distribution of contract deployments / addresses in the included GT sets on a time line (in million blocks).}
\label{fig:distribution_addr}
\end{minipage}

\medskip

\ITEM{Deployment blocks and forks.}
To put the addresses into a temporal context, we depict the deployment block in figure~\ref{fig:distribution_addr}.
We count the deployments in bins of 100\,000 blocks and depict them on a timeline of blocks (ticks per million blocks).

The latest block in the GT sets is 8\,M, while by the end of 2022, the main chain was beyond block 16.3\,M.
The deployment block also indicates, which EVM opcodes (introduced by a regular fork) were available.\footnote{%
An important opcode change occurred at block 7.28\,M with the introduction of the shift operations, which now appear in most contracts, and \op{create2}. At block 9.069\,M, \op{selfbalance} and \op{chainid} got introduced, and at block 12.9\,M \op{basefee}.
}
This information may be critical if a detection tool was developed before new opcodes were introduced.

\section{Consolidated Ground Truth} \label{sec:consolidation}
In this section, we describe the consolidation of the unified GT set.
It consists of (i) identifying entries pertaining to the same contract, (ii) marking conflicts within sets, (iii) mapping all assessments to a common taxonomy, (iv) determining the overlaps between the included sets, and (v) analyzing disagreements between the sets.

\subsection{Matching Contracts}\label{ssec:matching}

To detect assessments referring to the same contract, we match the address and the fingerprints of the codes (cf.\ section~\ref{ssec:completion}) according to the following considerations:
\begin{itemize}[topsep=0pt, itemsep=0pt]
\item Same address and chain means same contract, since none of the contracts in the sets was deployed via \textsc{CREATE2}.

\item Most assessments are based on the Solidity source code.
As the source usually specifies the admissible compiler versions (except when the missing directive is actually the weakness), the semantics of the program is fixed.
So, if two source codes have identical fingerprints and the names of the contracts under consideration are the same, the assessments refer to the same contract.

\item Assessments referring to deployment bytecodes with the same fingerprint can be considered as assessing the same contract, unless the checked property is tied to Solidity (like inheritance issues).
For the SWC classes, table~\ref{tab:coverage} indicates the visibility of the weakness by a checkmark in the last column.

\item Assessments referring to runtime codes with the same fingerprint are comparable only if the checked property is guaranteed to be detectable in this part of the code.
Typically, this holds for weaknesses depending on the contract being called by an adversary.
For the SWC classes, table~\ref{tab:coverage} indicates the visibility of the weakness in runtime code by a non-parenthesized checkmark in the last column.
\end{itemize}

\begin{table}[!ht]
  \centering
  \caption{Coverage of SWC Classes in the Consolidated GT Set.}
  \label{tab:coverage}
  \setlength\tabcolsep{3pt}
  \begin{tabular}{rr|rrl|c}
    \toprule
    SWC-id & \#Sets & pos. & neg. & Weakness & Bytecode \\
    \midrule
    100 & 1 &   1 &   1 & Function Default Visibility &  \\
    \rowcolor{grey1}
    101 & 7 & 807 & 322 & Integer Overflow and Underflow & (\YES) \\
    102 & 1 &   1 &   0 & Outdated Compiler Version &  \\
    103 & 3 & 513 &  46 & Floating Pragma &  \\
    \rowcolor{grey1}
    104 &10 & 426 &1402 & Unchecked Call Return Value & (\YES) \\
    105 & 4 &  62 &   5 & Unprotected Ether Withdrawal & \YES \\
    106 & 3 &   7 &   3 & Unprotected SELFDESTRUCT & \YES \\
    \rowcolor{grey1}
    107 &12 & 354 &2087 & Reentrancy & \YES \\
    108 & 2 &   3 &   1 & State Variable Default Visibility &  \\
    109 & 3 &   6 &   3 & Uninitialized Storage Pointer &  \\
    110 & 2 &  15 &   8 & Assert Violation & (\YES) \\
    111 & 2 &   2 &   2 & Use of Deprecated Solidity Functions &  \\
    112 & 4 &  33 &   6 & Delegatecall to Untrusted Callee & (\YES) \\
    \rowcolor{grey1}
    113 & 8 & 331 &1359 & DoS with Failed Call & (\YES) \\
    \rowcolor{grey1}
    114 & 8 & 535 & 726 & Transaction Order Dependence & \YES \\
    \rowcolor{grey1}
    115 & 7 &  79 &1619 & Authorization through tx.origin & (\YES) \\
    116 & 5 & 174 &   1 & Block values as a proxy for time & (\YES) \\
    117 & 2 &   3 &   2 & Signature Malleability & (\YES) \\
    118 & 4 &   9 &   3 & Incorrect Constructor Name &  \\
    119 & 3 &   4 &   3 & Shadowing State Variables &  \\
    \rowcolor{grey1}
    120 & 8 & 315 &1423 & Weak Sources of Randomness & (\YES) \\
    123 & 1 &   1 &   1 & Requirement Violation & (\YES) \\
    124 & 4 &  10 &   4 & Write to Arbitrary Storage Location & \YES \\
    125 & 2 &   2 &   2 & Incorrect Inheritance Order &  \\
    127 & 2 &   2 &   1 & Arbitrary Jump & \YES \\
    128 & 5 &  25 & 529 & DoS With Block Gas Limit & (\YES) \\
    129 & 2 &   4 &   1 & Typographical Error &  \\
    130 & 2 &   2 &   1 & Right-To-Left-Override control character &  \\
    131 & 1 &   2 &   2 & Presence of unused variables & (\YES) \\
    132 & 5 &  16 & 566 & Unexpected Ether balance & (\YES) \\
    133 & 2 &   2 &   3 & Hash Collisions & (\YES) \\
    134 & 2 &  18 &   0 & Message call with hardcoded gas amount & (\YES) \\
    135 & 2 &  12 & 525 & Code With No Effects & (\YES) \\
    136 & 2 &   3 &   3 & Unencrypted Private Data On-Chain &  \\
    \midrule
    995 & 2 &   2 &   1 & Short Address Attack & \YES \\
    996 & 3 &  51 &   1 & Honey Pot & \YES \\
    997 & 3 &  86 & 530 & Locked Ether & (\YES) \\
    999 & 1 &   3 &   7 & Other Arithmetic Issue & (\YES) \\
    \midrule
    (\YES) & \multicolumn{5}{l}{provided the weakness does not occur exclusively in the constructor}\\
    \bottomrule
  \end{tabular}
\end{table}

\subsection{Assessments Excluded from the Consolidated Set}\label{ssec:ignored}

For obvious reasons, we ignore assessments where either the judgment is n/a, or where the object of the assessment is ill defined.
The first condition affects 397 assessments, mostly from the set CodeSmells.
The second one eliminates 153 assessments from the Zeus set, as some contract identifiers do not allow us to extract a valid chain address, and the set does not provide any further information.

It is well known that contracts like wallets or tokens have been deployed identically numerous times.
Often, this fact is not taken into account when collecting contract samples, such that the same contract may end up in a set multiple times, albeit under different addresses.
Therefore, we check the sets for multiple assessments of the same code, to find contradictions and duplicates.

Surprisingly, the Zeus set contains 18 contradictions already on the level of its own identifiers (meaning that the same identifier is listed multiple times, with diverging assessments) and 30 more when applying the criteria laid out in the last section.
Moreover, we find 103 conflicts in the set CodeSmells, 6 in Doublade, and 3 in JiuZhou.
These assessments are excluded from the consolidated set.

For duplicates, all but one assessment are redundant and can be ignored.
We find duplicates in almost every set (the number in parentheses gives the ignored assessments): CodeSmells (853), ContactFuzzer (4), Doublade (34), eThor (6), EthRacer (5), EverEvolvingGame (52), NPChecker (31), SBcurated (16), SolidiFI (7), SWCregistry (1), and Zeus (3009).

For a summary of the exclusions, see the column `ignored' in table~\ref{tab:gt_included}.
Table~\ref{tab:ignored} in the appendix gives a breakdown of the ignored assessments by reason.

\subsection{Mapping of Individual Assessments to a Common Taxonomy}\label{ssec:mapping}

In order to compare assessments in different sets, we map the properties of each set to a corresponding class in a suitable taxonomy.
The SWC registry\footnote{\url{https://swcregistry.io}} provides such a widely used taxonomy with 37 weakness classes.
Each of its classes has a numeral identifier, weakness title, CWE parent and some code samples.
See table~\ref{tab:mapping} appendix~\ref{sec:mappings} for the details of the mapping we applied.

\subsubsection{Coverage of SWC classes.}
Table~\ref{tab:coverage} shows how well the SWC classes are covered by positive and negative assessments, and how many sets contribute assessments to the class.
Popular weaknesses with seven or more contributing GT sets are marked in gray.
At the bottom, we add the classes 995--999 to account for weaknesses missing from the SWC registry.

Even after combining all GT sets into a unified ground truth, the coverage of the SWC classes remains highly uneven.
This can be attributed to the intention behind most benchmark sets: to support the test of tools for automated vulnerability detection.
And tools aim for ``interesting'' weaknesses.

All weaknesses can be detected in the source code.
The last column of table~\ref{tab:coverage} marks those that are also detectable in the bytecode.
A checkmark in parentheses indicates that the weakness may occur in the constructor and thus is not necessarily visible in the deployed runtime code.

\subsubsection{Comparison of Weaknesses.}
It is intrinsically difficult to compare weaknesses across GT sets due to 
(i) vague or missing \textit{definitions} of weaknesses or vulnerabilities by the authors,
(ii) the unclear \textit{relationship} between definitions,
(iii) the ambiguous \textit{mapping} of a weakness to a corresponding class, and
(iv) the uncertain \textit{structuring} of weaknesses according to reasonable aspects.
The authors of a GT set use definitions that are rarely a perfect match for a taxonomy.
Thus, when comparing weaknesses via a taxonomy, disagreements have to be checked for definition mismatches.

\subsection{Overlaps}
In order to find disagreements between the cleaned GT sets, we first determine their overlap.
For each pair of GT sets, table~\ref{tab:overlap_assessment} gives the number of non-ignored assessments that map to the same SWC class.
The diagonal gives the total number of mapped assessments in a GT set.
The upper-left block relates to the group of \emph{wild} sets, while the lower-right block concerns the \emph{crafted} ones.
SBcurated is a special case as it is a mixture of \emph{crafted} and \emph{wild} contracts, with several of its \emph{crafted} contracts taken from the SWCregistry.
\begin{table}
  \caption{Overlap of Mapped Assessments in the Consolidated GT Set.}%
  \label{tab:overlap_assessment}%
  \centering
  \setlength\tabcolsep{2pt}%
  \newcommand\OL{\cellcolor{grey1}}%
  \newcommand\OO{\color{grey2}{0}}%
  \begin{tabular}{@{}l|rrrrrrrr|rrrrr@{}}
    \toprule
    Set & \SW{CodeSmells} & \SW{ContractFuzzer} & \SW{Doublade} & \SW{eThor} & \SW{EthRacer} & \SW{EverEvolvingG} & \SW{NPChecker} & \SW{Zeus} & \SW{JiuZhou} & \SW{NotSoSmartC} & \SW{SBcurated} & \SW{SolidiFI} & \SW{SWCregistry}  \\
    \midrule
    CodeSmells     &   5300 & \OL  6 & \OL 4 & \OL  26 & \OO & \OO & \OL 14 & \OL 145 & \OO & \OL1& \OO & \OO & \OO \\
    ContractFuzzer & \OL  6 &    375 &     6 & \OO     & \OO & \OO & \OL  3 & \OL  15 & \OO & \OO &  10 & \OO & \OO \\
    Doublade       & \OL  4 &      6 &   279 & \OL   2 & \OO & \OO &      2 & \OL  10 & \OO & \OO &   7 & \OO & \OO \\
    eThor          & \OL 26 & \OO    & \OL 2 &     702 & \OO & \OO & \OL 25 & \OL 691 & \OO & \OO & \OO & \OO & \OO \\
    EthRacer       & \OO    & \OO    & \OO   & \OO     & 111 & \OO & \OO    & \OL   5 & \OO & \OO & \OO & \OO & \OO \\
    EverEvolvingG. & \OO    & \OO    & \OO   & \OO     & \OO &  65 & \OO    & \OO     & \OO & \OO & \OO & \OO & \OO \\
    NPChecker      & \OL 14 & \OL  3 &     2 & \OL  25 & \OO & \OO &    219 & \OL 128 & \OO & \OO & \OO & \OO & \OO \\
    Zeus           & \OL145 & \OL 15 & \OL10 & \OL 691 &\OL5 & \OO & \OL128 &    7323 & \OO & \OO &   1 & \OO & \OO \\
    \midrule
    JiuZhou        & \OO    & \OO    & \OO   & \OO     & \OO & \OO & \OO    & \OO     & 129 & \OO & \OO & \OO &   2 \\
    NotSoSmartC    & \OL  1 & \OO    & \OO   & \OO     & \OO & \OO & \OO    & \OO     & \OO &  32 &   7 & \OO & \OO \\
    SBcurated      & \OO    &     10 &     7 & \OO     & \OO & \OO & \OO    &       1 & \OO &   7 & 129 & \OO &  31 \\
    SolidiFI       & \OO    & \OO    & \OO   & \OO     & \OO & \OO & \OO    & \OO     & \OO & \OO & \OO & 343 & \OO \\
    SWCregistry    & \OO    & \OO    & \OO   & \OO     & \OO & \OO & \OO    & \OO     &   2 & \OO &  31 & \OO & 116 \\
    \bottomrule
  \end{tabular}
\end{table}

Of the total of 20\,498 cleaned assessments, 18\,409 appear in only one of the sets, while 2\,089 have overlaps in one or more sets.
As expected, there is more overlap within the \emph{wild} group than the \emph{crafted} group or between the groups.

\subsection{Disagreements between the Sets}
In table~\ref{tab:overlap_assessment}, overlaps containing disagreements are marked in gray.
Of the 2\,098 overlapping assessments, 458 disagree with another one, involving eight GT sets and six SWC classes as listed in table~\ref{tab:disagree}.

The disagreements constitute an interesting area of investigation.
While some disagreements are due to diverging definitions of weaknesses that were mapped to the same SWC class, quite a few turn out to be actual inconsistencies under the authors' original definitions.
Table~\ref{tab:disagree} summarizes the results of our manual evaluation.
For each affected set, it gives the total number of assessments that disagree with an assessment of another set as well as a breakdown by SWC class.
A table entry is marked red if our evaluation revealed assessment errors, giving also the number of such errors.
\begin{table}
  \newcommand\da[2]{\WA #1 of #2}
  \centering
  \caption{Number of Disagreements in the Unified GT Set, with the Errors.}
  \label{tab:disagree}
  \begin{tabular}{lr@{ \ }|@{ \ }c@{\quad}c@{\quad}c@{\quad}c@{\quad}c@{\quad}c}
    \toprule
    Dataset & total & 104 & 107 & 113 & 114 & 120 & 997 \\
    \midrule
    CodeSmells & 34 & 8 & \da{6}{9} & 8 &: & \da{7}{7} & \da{2}{2} \\
    ContractFuzzer & 13 & 7 & : &: &: & \da{4}{4} & 2 \\
    Doublade & 6 & 3 & \da{3}{3} &: &: &: &: \\
    eThor & 166 & : & 166 & : & : & : & : \\
    EthRacer & 2 & : & : & : & \da{2}{2} & : & :\\
    NotSoSmartC & 1 &: & 1 &: &: &: &: \\
    NPChecker & 25 & \da{3}{6} & \da{1}{7} & 7 & \da{1}{2} & \da{2}{3} & : \\
    Zeus & 211 & 18 & \da{3}{163} & 15 & \da{1}{4} & 11 & : \\
    \bottomrule
  \end{tabular}
\end{table}

Since reentrancy is the most popular weakness, it appears in 12 GT sets and gives rise to most overlaps: of the 2\,098 overlapping assessments, 1\,480 pertain to reentrancy (SWC 107).
Thus, it is not surprising that we observe the highest number of disagreements (182) and errors (13) for reentrancy.

With 42 disagreements, SWC 104 is second in frequency.
However, we identified only three errors, with the other disagreements resulting from diverging definitions.
While SWC 113 shows no errors, for SWC 114, 120 and 997 on average about half the disagreements are errors.

\subsection{Random Inspection}

To gain further insights into the quality of the assessments, we randomly select 80 assessments from the consolidated ground truth, in order to manually check them.
Table~\ref{tab:checked} shows, for each GT set and SWC class, the number of checked assessments as well as the number of errors.
\begin{table}
  \newcommand\da[2]{\WA #1 of #2}
  \centering
  \caption{Number of Manually Checked Assessments, with the Errors.}
  \label{tab:checked}
   \begin{tabular}{lc@{\quad}c@{\quad}c@{\quad}c@{\quad}c@{\quad}c@{\quad}c@{\quad}c@{\quad}c}
    \toprule
    Set~~\textbackslash~~SWC & 101 & 104 & 107 & 112 & 113 & 114 & 115 & 120 & 997 \\
    \midrule
    CodeSmells     & :       &  1      & 3 & :        & 3 & : & : & : &\da{2}{7}\\
    ContractFuzzer & :       & 10      & 3 &\da{4}{11}& : & : & : & 4 & 9       \\
    Doublade       & :       &  1      & 2 & :        & : & : & 1 & : & :       \\
    NPChecker      & :       &\da{1}{4}& : & :        & : & : & : & : & :       \\
    SBcurated      & :       &  1      & 2 & :        & : & : & : & : & :       \\
    Zeus           &\da{4}{6}&  7      & : & :        & 1 & 4 & : & : & :       \\
    \bottomrule
  \end{tabular}
\end{table}

\section{Discussion} \label{sec:discussion}
\subsection{Data Quality} \label{ssec:quality}
In table~\ref{tab:data_quality}, we assess the data quality of the GT sets along the dimensions proposed by Bosu et al.~\cite{Bosu2013taxonomy}.
The mapping of ranges to the three symbols is specified in table~\ref{tab:criteria} in the appendix.

\begin{table}
\caption{Data Quality of Ground Truth Sets.}\label{tab:data_quality}
\centering
\setlength\tabcolsep{3pt}
\begin{tabular}{lr|cccccccc|ccccc|c}
\toprule
&&\multicolumn{8}{c|}{wild}&\multicolumn{5}{c|}{crafted}& \\
\multicolumn{2}{r|}{Quality Aspect\quad~\SW{Set}}
                & \SW{CodeSmells}
                       & \SW{ContractFuzzer}
                              & \SW{Doublade}
                                     & \SW{eThor}
                                            & \SW{EthRacer}
                                                   & \SW{EverEvolvingG}
                                                          & \SW{NPChecker}
                                                                 & \SW{Zeus}
                                                                        & \SW{JiuZhou}
                                                                               & \SW{NotSoSmartC}
                                                                                      & \SW{SBcurated}
                                                                                             & \SW{SolidiFI}
                                                                                                    & \SW{SWCregistry}
                                                                                                           & \SW{Consolidated GT} \\
\midrule
\multirow{3}{*}{Accuracy} 
& completeness  & \bad & \good& \med & \good& \bad & \bad & \bad & \bad & \med & \med & \med & \med & \good& \good\\
& irreduncancy  & \med & \good& \med & \good& \med & \bad & \bad & \bad & \good& \good& \good& \med & \good& \good\\
& consistency   & \bad & \med & \bad & \good& \med & \good& \med & \bad & \bad & \good& \good& \good& \good& \med \\
\midrule
\multirow{3}{*}{Relevance} 
& heterogeneity	& \bad & \bad & \bad & \bad & \bad & \bad & \bad & \bad & \med & \bad & \med & \bad & \med & \med \\
& data quantity & \good& \med & \med & \med & \bad & \med & \med & \good& \med & \med & \med & \med & \med & \good\\
& timeliness    & \bad & \bad & \bad & \bad & \bad & \bad & \bad & \bad & \bad & \bad & \bad & \bad & \bad & \bad \\
\bottomrule
\end{tabular}
\end{table}

\subsubsection{Accuracy.}
All sets provide a minimum of data, but we had to complete the data of about two thirds of the sets.
Redundancy exists in many \textit{wild} GT sets, to varying degrees.
The main concern regards inconsistencies~-- the key aspect of a GT~-- which we encountered in six \textit{wild} GT sets.
We improved the data quality (i)~by data completion, (ii)~by eliminating redundant and contradictory assessments within sets, and (iii)~by resolving disagreements between sets.
Thus, we could increase the accuracy in the consolidated GT set in all aspects.
However, random inspections revealed further inconsistencies; the overall accuracy would benefit from further checks.

\subsubsection{Relevance.}
The GT sets mostly lack heterogeneity, often provide a smallish amount of data, and above all lack recent data.
By merging 13 GT sets, we could improve the amount of data (number of positive and especially negative examples) and some aspects of heterogeneity, like the number of weaknesses and weakness classes.
However, there is still a bias with respect to language, time, contracts submitted to Etherscan, and popular vulnerabilities.

\subsubsection{Bias.} We identify various sources contributing to a bias in the GT sets. \emph{Contract source:} the verified source codes from Etherscan have been deliberately submitted as open source. \emph{Language:} the overwhelming majority is written in Solidity. \emph{Origin:} deployed bytecodes show less variety than sets with vulnerabilities injected systematically. \emph{Creation time:} most GT sets mainly contain Solidity source code from 2016 to 2018.

\subsection{Related Work}

Each of the thirteen original datasets can be regarded as distantly related work; see section~\ref{sec:benchmarks} for a description.
Concerning the construction of a unified GT set, we only find AutoMESC~\cite{Soud2022automesc}.
In this work, Soud et al.\  choose five source code datasets \cite{Yashavant2022,Durieux2020,Ren2021,Zhang2020framework,Grech2019} that address 10 vulnerabilities, which are detected by one or more of the tools HoneyBadger, Mythril, Maian, Osiris, Slither, SmartCheck, Solhint.
They apply as inclusion criteria: recent (up to three years old), public, Ethereum, corresponding publication or GitHub repo;
and exclude commercial and competition datasets, and sets that just provide one example per vulnerability.
For unification, they use a file-ID per contract (without checking for non-obvious duplicates).
For consolidation (identifying duplicates, mapping the assessments to a common taxonomy, and resolving contradictory assessments), they discard the original classification, and replace it with a simple majority vote of the seven selected tools if those claim to detect the weakness (after mapping the tool findings to a common taxonomy).
They claim that there is neither redundancy nor inconsistency in the five datasets included.

\subsection{Challenges in Identifying Weaknesses}

\ITEM{Ambiguous definitions of weaknesses.}
Hardly any weakness possesses a commonly accepted, precise definition.
As a consequence, seemingly contradictory assessments of a contract by different datasets may actually result from applying subtly different definitions.

\ITEM{Weakness vs.\ vulnerability.}
There is no agreement among dataset authors whether to aim for exploitable or potential issues.

\ITEM{Intended purpose.}
The verdict on whether a weakness is considered a vulnerability also depends on the purpose of a contract.
An apparent weakness may be actually the intended behavior of the contract (e.g.\ a faucet that ``leaks'' Ether).

\ITEM{Contracts in isolation.}
The included datasets consider single-contract weaknesses only (discounting the attack contract).
However, vulnerabilities may be the result of several interacting contracts.
A single contract may not provide sufficient context to be classified as vulnerable on its own.

\subsection{Reservations about Majority Voting}
Due to the scarcity of GT data, some authors resort to pseudo-GT data.
They run several vulnerability detection tools on selected contracts and obtain the judgment by comparing the number of positive results to a threshold.
This approach is debatable for the following reasons.

\ITEM{Weakness vs.\ vulnerability.}
Most tools detect code patterns that indicate a weakness, regardless of whether it can be actually exploited.
Hence, false positives (and, to a lesser extent, false negatives) are rather the norm than the exception.
Thus, majority vote may turn false positives into a positive assessment.
\ITEM{Tool genealogy.}
Tools form families by being derived from common ancestors (like Oyente), by implementing the same approach (like symbolic execution, taint analysis, or fuzzing), or by relying on the same basic components (like GigaHorse, Rattle, Z3, or Soufflé).
Related tools may misjudge a contract in a similar way and outnumber tools with the correct result.
\ITEM{Diverging definitions of weaknesses.}
Even if labeled the same, the weaknesses detected by any two tools are not quite the same.
Rather, we are faced with tools voting on a weakness that is more or less similar to what they can detect.

\section{Conclusions}
Publicly available ground truth data for smart contract weaknesses is scarce, but much needed. 
Our consolidated ground truth is an appreciation of the commendable efforts by others and hopefully renders the included GT sets more usable to the community.
The consolidated ground truth described in this paper is available from \url{http://github.com/gsalzer/cgt}

\ITEM{Data Quality.} \textbf{Accuracy}.
Some GT sets were compiled more diligently than others, so our consolidation increased the quality by reducing redundancy and applying further consistency and quality checks.
Most GT sets contain basic data only, which we completed with the help of external databases.
\textbf{Relevance}.
As most GT sets are neither heterogeneous nor balanced by themselves, we could achieve an improvement by merging them.
However, the size still needs to be increased, and more recent data is desirable.

\ITEM{Future Work.} \textbf{Granularity.}
This unified and consolidated GT set is constructed on contract level.
Information on the location of weaknesses within the contracts, like the line number in the source or the offset in the byte code, is available only for two small datasets, and was omitted here.
\textbf{Severity Level}.
Assigning a severity level to a weakness would further improve the GT set, but is a difficult topic on its own.
\textbf{Updates} are important.
We invite everyone to contribute by adding GT collections, taxonomies, levels of granularity or severity, proofs and exploits.

\bibliographystyle{splncs04}
\bibliography{references}

\begin{thebibliography}{10}
\providecommand{\url}[1]{\texttt{#1}}
\providecommand{\urlprefix}{URL }
\providecommand{\doi}[1]{https://doi.org/#1}

\bibitem{Ashouri2020}
Ashouri, M.: {Etherolic: A practical security analyzer for smart contracts}.
  Proceedings of the ACM Symposium on Applied Computing pp. 353--356 (2020).
  \doi{10.1145/3341105.3374226}

\bibitem{Bosu2013taxonomy}
Bosu, M.F., MacDonell, S.G.: A taxonomy of data quality challenges in empirical
  software engineering. In: 2013 22nd Australian Software Engineering
  Conference. pp. 97--106. IEEE (2013). \doi{10.1109/ASWEC.2013.21}

\bibitem{Chen2020defining}
Chen, J., Xia, X., Lo, D., Grundy, J., Luo, X., Chen, T.: Defining smart
  contract defects on ethereum. IEEE Transactions on Software Engineering
  pp.~1--1 (2020). \doi{10.1109/TSE.2020.2989002}

\bibitem{Chen2021DefectChecker}
Chen, J., Xia, X., Lo, D., Grundy, J., Luo, X., Chen, T.: Defectchecker:
  Automated smart contract defect detection by analyzing evm bytecode. IEEE
  Transactions on Software Engineering  (2021). \doi{10.1109/TSE.2021.3054928}

\bibitem{Dias2021}
Dias, B., Ivaki, N., Laranjeiro, N.: An empirical evaluation of the
  effectiveness of smart contract verification tools. In: 2021 IEEE 26th
  Pacific Rim International Symposium on Dependable Computing (PRDC). pp.
  17--26 (2021). \doi{10.1109/PRDC53464.2021.00013}

\bibitem{Dika2018}
Dika, A., Nowostawski, M.: Security vulnerabilities in ethereum smart
  contracts. In: 2018 IEEE International Conference on Internet of Things
  (iThings) and IEEE Green Computing and Communications (GreenCom) and IEEE
  Cyber, Physical and Social Computing (CPSCom) and IEEE Smart Data
  (SmartData). pp. 955--962. IEEE (2018).
  \doi{10.1109/Cybermatics_2018.2018.00182}

\bibitem{Durieux2020}
Durieux, T., Ferreira, J.F., Abreu, R., Cruz, P.: {Empirical review of
  automated analysis tools on 47,587 Ethereum smart contracts}. In: Proceedings
  of the ACM/IEEE 42nd International Conference on Software Engineering. pp.
  530--541. ACM, New York, NY, USA (2020). \doi{10.1145/3377811.3380364}

\bibitem{Ferreira2020}
Ferreira, J.F., Cruz, P., Durieux, T., Abreu, R.: {SmartBugs: A Framework to
  Analyze Solidity Smart Contracts}. In: 35th IEEE/ACM International Conference
  on Automated Software Engineering (ASE 2020). pp. 1349--1352. ACM (2020).
  \doi{10.1145/3324884.3415298}

\bibitem{Ghaleb2020SolidiFI}
Ghaleb, A., Pattabiraman, K.: How effective are smart contract analysis tools?
  evaluating smart contract static analysis tools using bug injection. In:
  Proceedings of the 29th ACM SIGSOFT International Symposium on Software
  Testing and Analysis. pp. 415--427. ISSTA 2020, Association for Computing
  Machinery (2020). \doi{10.1145/3395363.3397385}

\bibitem{Grech2019}
Grech, N., Brent, L., Scholz, B., Smaragdakis, Y.: Gigahorse: thorough,
  declarative decompilation of smart contracts. In: 2019 IEEE/ACM 41st
  International Conference on Software Engineering (ICSE). pp. 1176--1186. IEEE
  (2019). \doi{10.1109/ICSE.2019.00120}

\bibitem{gupta2019analysis}
Gupta, B.C.: Analysis of Ethereum Smart Contracts - A Security Perspective.
  Master's thesis, Department of Computer Science and Engineering, Indian
  Institute of Technology Kanpur (2019),
  \url{https://security.cse.iitk.ac.in/sites/default/files/17111011.pdf}

\bibitem{Gupta2020Insecurity}
Gupta, B.C., Kumar, N., Handa, A., Shukla, S.K.: {An Insecurity Study of
  Ethereum Smart Contracts}. In: Security, Privacy, and Applied Cryptography
  Engineering (SPACE 2020), vol. LNCS 12586, pp. 188--207. Springer
  International Publishing (2020). \doi{10.1007/978-3-030-66626-2_10}

\bibitem{Ji2021}
Ji, S., Kim, D., Im, H.: {Evaluating Countermeasures for Verifying the
  Integrity of Ethereum Smart Contract Applications}. IEEE Access  \textbf{9},
  90029--90042 (2021). \doi{10.1109/access.2021.3091317}

\bibitem{Jiang2018ContractFuzzer}
Jiang, B., Liu, Y., Chan, W.K.: Contractfuzzer: Fuzzing smart contracts for
  vulnerability detection. In: Proceedings of the 33rd ACM/IEEE International
  Conference on Automated Software Engineering. pp. 259--269. ASE 2018,
  Association for Computing Machinery (2018). \doi{10.1145/3238147.3238177}

\bibitem{Kalra2018}
Kalra, S., Goel, S., Dhawan, M., Sharma, S.: {ZEUS: Analyzing Safety of Smart
  Contracts}. In: NDSS Symposion. NDSS, Internet Society (2018).
  \doi{10.14722/ndss.2018.23082}

\bibitem{Kim2020}
Kim, K.B., Lee, J.: {Automated Generation of Test Cases for Smart Contract
  Security Analyzers}. IEEE Access  (2020). \doi{10.1109/ACCESS.2020.3039990}

\bibitem{Kolluri2019}
Kolluri, A., Nikolic, I., Sergey, I., Hobor, A., Saxena, P.: Exploiting the
  laws of order in smart contracts. In: Proceedings of the 28th ACM SIGSOFT
  International Symposium on Software Testing and Analysis. pp. 363--373. ISSTA
  2019, Association for Computing Machinery, New York, NY, USA (2019).
  \doi{10.1145/3293882.3330560}

\bibitem{Perez2021}
Perez, D., Livshits, B.: Smart contract vulnerabilities: Vulnerable does not
  imply exploited. In: 30th $\{$USENIX$\}$ Security Symposium ($\{$USENIX$\}$
  Security 21) (2021),
  \url{https://www.usenix.org/system/files/sec21-perez.pdf}

\bibitem{Rameder2022}
Rameder, H., Angelo, M.D., Salzer, G.: Review of automated vulnerability
  analysis of smart contracts on ethereum. Front. Blockchain - Smart Contracts
  (2022). \doi{10.3389/fbloc.2022.814977}

\bibitem{Ren2021}
Ren, M., Yin, Z., Ma, F., Xu, Z., Jiang, Y., Sun, C., Li, H., Cai, Y.:
  Empirical evaluation of smart contract testing: what is the best choice? In:
  Proceedings of the 30th ACM SIGSOFT International Symposium on Software
  Testing and Analysis. pp. 566--579 (2021). \doi{10.1145/3460319.3464837}

\bibitem{Schneidewind2020}
Schneidewind, C., Grishchenko, I., Scherer, M., Maffei, M.: {EThor: Practical
  and Provably Sound Static Analysis of Ethereum Smart Contracts}. In:
  Proceedings of the ACM Conference on Computer and Communications Security.
  pp. 621--640 (may 2020). \doi{10.1145/3372297.3417250}

\bibitem{Soud2022automesc}
Soud, M., Qasse, I., Liebel, G., Hamdaqa, M.: Automesc: Automatic framework for
  mining and classifying ethereum smart contract vulnerabilities and their
  fixes. arXiv preprint arXiv:2212.10660  (2022).
  \doi{10.48550/arXiv.2212.10660}

\bibitem{FerreiraTorres2021confuzzius}
Torres, C.F., Iannillo, A.K., Gervais, A., State, R.: Confuzzius: A data
  dependency-aware hybrid fuzzer for smart contracts. In: 2021 IEEE European
  Symposium on Security and Privacy (EuroS\&P). pp. 103--119 (2021).
  \doi{10.1109/EuroSP51992.2021.00018}

\bibitem{Wang2019detecting}
Wang, S., Zhang, C., Su, Z.: Detecting nondeterministic payment bugs in
  ethereum smart contracts. Proceedings of the ACM on Programming Languages
  (PACMPL)  \textbf{3}(189),  1--29 (2019). \doi{10.1145/3360615}

\bibitem{Xue2019doublade}
Xue, Y., Ye, J., Ma, M., Ma, L., Li, Y., Wang, H., Lin, Y., Peng, T., Liu, Y.:
  Doublade: Unknown vulnerability detection in smart contracts via abstract
  signature matching and refined detection rules. arXiv preprint
  arXiv:1912.04466  (2019). \doi{10.48550/arXiv.1912.04466}

\bibitem{Yashavant2022}
Yashavant, C.S., Kumar, S., Karkare, A.: Scrawld: A dataset of real world
  ethereum smart contracts labelled with vulnerabilities. arXiv preprint
  arXiv:2202.11409  (2022). \doi{10.48550/arXiv.2202.11409}

\bibitem{Zhang2020framework}
Zhang, P., Xiao, F., Luo, X.: A framework and dataset for bugs in ethereum
  smart contracts. In: IEEE International Conference on Software Maintenance
  and Evolution (ICSME). pp. 139--150. ICSME 2020, IEEE (2020).
  \doi{10.1109/icsme46990.2020.00023}

\bibitem{Zhou2020}
Zhou, S., Yang, Z., Xiang, J., Cao, Y., Yang, Z., Zhang, Y.: An ever-evolving
  game: Evaluation of real-world attacks and defenses in ethereum ecosystem.
  In: 29th {USENIX} Security Symposium ({USENIX} Security 20). pp. 2793--2810.
  USENIX Security 2020, {USENIX} Association (2020),
  \url{https://www.usenix.org/conference/usenixsecurity20/presentation/zhou-shunfan}

\end{thebibliography}

\clearpage
\appendix
\clearpage
\section{GT Sets not Included} \label{sec:excluded}
Sets that do not assess contracts on their own, but rather reuse sets from others were not included and are mentioned in this section.

\subsection{Sets Contained in Other Collections}
Durieux et al.~\cite{Durieux2020} used two sets for a tool review: 47\,518 verified contracts and 69 annotated vulnerable smart contracts. They are both contained in SmartBugs~\cite{Ferreira2020}.

Dika et al.~\cite{Dika2018} use data from other benchmark sets, but do not provide theirs for download.

Ferreira Torres et al.~\cite{FerreiraTorres2021confuzzius} use a set of 128 contracts adapted from SmartBugs and verified contracts from Etherscan.io to evaluate their tool ConFuzzius.

Dias et al.~\cite{Dias2021} use a set of 222 contracts from different sources (e.g.\ swcregistry.io, \cite{Zhang2020framework}), of which  94 are vulnerable, 30 misleading, and 98 fixed contracts.

Zhou et al.~\cite{Zhou2020} ``contacted the authors of 11 prior works ... and obtained eight replies and six datasets''.
We included it since the sets were re-assessed.

Ren et al.~\cite{Ren2021} collected contracts from SolidiFI~\cite{Ghaleb2020SolidiFI}, CVE.mitre.org, Etherscan.io, and SWCregistry.io for their benchmark set. 

Chen et al.~\cite{Chen2021DefectChecker} used the GT set from CodeSmells~\cite{Chen2020defining} for the tool DefectChecker.

\subsection{Sets not Available or not a Ground Truth}
The set of~\cite{Ji2021} consists of ``237 benchmark codes'' and is not available, but Ji et al.\ list their sources as ``SWC registry (Smart Contract Weakness Classification and Test Cases), Smart-Bugs SB curated dataset, VeriSmart-benchmarks, Zeus dataset, and eThor dataset''.

Ashouri et al.~\cite{Ashouri2020} describe their set as ``a crafted benchmark suite, comprising several real-world and synthetic smart contracts along with 98 safety features'', but the set is not provided.

Gupta et al.~\cite{gupta2019analysis,Gupta2020Insecurity} list their sources as ``Smart Contract Weakness Classification (SWC) Registry, (Not So) Smart Contracts, EVM Analyzer Benchmark Suite, research papers, theses and books, various blog posts, articles, etc. ... A final set of 180 vulnerable contracts is assorted [...] Additionally real world Ethereum smart contracts were collected and distilled to a set of 2\,715 unique contracts, for which 2\,053 Solidity source code files could be retrieved'', but the set is not available.

For Zeus~\cite{Kalra2018}, the set is provided view-only and has unclear references to the assessed contracts (just contract names or addresses, but no (source) code).
We included it anyway since it was distributed among researchers.

Neither the tool TestBreeder~\cite{Kim2020} nor the test cases it generated are available.

Perez et al.~\cite{Perez2021} list their sources as ``A total set of 821.219 smart contracts used in the studies of 6 tools were collected from the respective authors: Oyente, Zeus, MAIAN, Securify, MadMax, and teEther. 23.327 contracts of this set were deemed vulnerable to at least one of the described vulnerabilities'', but the set is not available.

\clearpage
\section{Tables}
\begin{table}[!ht]
\centering 
\caption{Included GT Sets -- Contents.}\label{tab:gt_contents}
\setlength\tabcolsep{2pt}
\begin{tabular}{@{}lB{2.1cm}B{2.0cm}lcccc@{}}
\toprule
Set & contract \quad given by & assessments given by & \SW{Solidity} & \SW{verified contract} & \SW{address} & \SW{deployment code} & \SW{runtime bytecode} \\
\midrule
\href{https://github.com/Jiachi-Chen/TSE-ContractDefects}{CodeSmells} & address & \texttt{csv} &  &  & \YES &  &  \\
\rowcolor{grey1}
\href{https://github.com/gongbell/ContractFuzzer/tree/master/examples}{ContractFuzzer} & Solidity src & filepath & \YES & \YES &  & mostly & mostly  \\
\href{https://drive.google.com/file/d/1k0Edw2r1Z59WBc8SFbeh85hJMydGNPGz/view}{Doublade} & Solidity src, address & filepath, \texttt{csv} & \YES & \YES & \YES &  &  \\
\rowcolor{grey1}
\href{https://secpriv.wien/ethor/}{eThor} & address & \texttt{csv} & mostly & mostly & \YES &  & \YES \\
\href{https://drive.google.com/file/d/1190VXwu502M-vgT8yyuFp0lFUVlxnMhO/view?usp=sharing}{EthRacer} & address & \texttt{csv} &  &  & \YES &  &  \\
\rowcolor{grey1}
\href{https://drive.google.com/open?id=1xLssDxYWyKFCwS5HUrQaSex0uwJRSvDi}{EverEvolvingG} & address & \texttt{json} &  &  & \YES &  &  \\
\href{https://www.dropbox.com/sh/90tm5drmeep9bqy/AAB0jKxkIevNct2eIvsYb7Oqa?dl=0}{NPChecker} & address & \texttt{xlsx} &  &  & \YES &  &  \\
\rowcolor{grey1}
\href{https://goo.gl/kFNHy3}{Zeus} & string & \texttt{csv} &  &  & mostly &  &  \\
\midrule
\href{https://github.com/xf97/JiuZhou}{JiuZhou} & Solidity src & filepath, text & \YES &  &  & \YES &  \\
\rowcolor{grey1}
\href{https://github.com/crytic/not-so-smart-contracts/}{NotSoSmartC} & Solidity src & filepath, text & \YES & some & some &  &  \\
\href{https://github.com/smartbugs/smartbugs-curated}{SBcurated} & Solidity src & \texttt{json} & \YES & some & some &  &   \\
\rowcolor{grey1}
\href{https://github.com/DependableSystemsLab/SolidiFI-benchmark}{SolidiFI} & Solidity src & filepath & \YES &  &  &  & \\
\href{swcregistry.io}{SWCregistry} & Solidity src & \texttt{json}, \texttt{yaml} & \YES & some &  & \YES & \YES \\
\bottomrule
\end{tabular}
\end{table}

\begin{table}[!ht]
\centering 
\caption{Methods Used to Assess the Contracts.} \label{tab:gt_methods}
\begin{tabular}{lp{9.1cm}}
\toprule
Sets & Method of Assessment \\
\midrule
CodeSmells~\cite{Chen2020defining} & Two researchers assessed the source code and discussed disagreements. \\
\rowcolor{grey1}
ContractFuzzer~\cite{Jiang2018ContractFuzzer} & Manual verification of vulnerable SCs detected by own tool. \\
Doublade~\cite{Xue2019doublade} & Manual inspection of results by own tool to identify false positives. \\
\rowcolor{grey1}
eThor~\cite{Schneidewind2020} & Manually re-assessed GT of Zeus, including some bytecode-only SCs with less than 6000 bytes. \\
EthRacer~\cite{Kolluri2019} & From the contracts flagged by own tool, 127 were randomly selected and manually checked. \\
\rowcolor{grey1}
EverEvolvingG~\cite{Zhou2020} & Assessments are based on inspecting transactions rather than code. \\
NPChecker~\cite{Wang2019detecting} & 50 SCs with Solidity source code randomly selected and manually inspected. \\
\rowcolor{grey1}
Zeus~\cite{Kalra2018} & Manually validated each result of own tool. \\
\midrule
JiuZhou~\cite{Zhang2020framework} & SCs were collected (SWCregisrty, NotSoSmartC), modified or hand-crafted. \\
\rowcolor{grey1}
\href{https://github.com/crytic/not-so-smart-contracts/}{NotSoSmartC} & Collection of mishaps for educational purposes. \\
SBcurated~\cite{Ferreira2020} & SCs were picked from other collections and manually assessed. \\
\rowcolor{grey1}
SolifiFI~\cite{Ghaleb2020SolidiFI} & Generated from 50 Solidity SCs by injecting 7 types of bugs. \\
\href{swcregistry.io}{SWCregistry} & Collection of examples for known vulnerabilities, manually verified. \\
\bottomrule
\end{tabular}
\end{table}
\begin{table}
\centering
\caption{GT Assessments with Issues (set to `ignore').} \label{tab:ignored}
\setlength\tabcolsep{2pt}
\begin{tabular}{lrrrrrrrrrrrrr}
\toprule
Issue reason & \SW{CodeSmells} &\SW{ContractFuzzer} & \SW{Doublade} & \SW{eThor}& \SW{EthRacer} & \SW{EverEvolvingG.} & \SW{JiuZhou} & \SW{NotSoSmartC} & \SW{NPChecker} & \SW{SBcurated} & \SW{SolidiFI} & \SW{SWCregistry} & \SW{Zeus} \\
  \midrule
invalid contract reference          &     0 & 0 &   0 &  0 &  0 &  0 &  0 &  0 &  0 &  0 &  0 & 0 &   153 \\
status is na                        &   374 & 0 &   0 & 12 & 11 &  0 &  0 &  0 &  0 &  0 &  0 & 0 &     0 \\
\midrule
contradiction for id                &     0 & 0 &   0 &  0 &  0 &  0 &  0 &  0 &  0 &  0 &  0 & 0 & \WA18 \\
conflict for address                &     0 & 0 &   0 &  0 &  0 &  0 &  0 &  0 &  0 &  0 &  0 & 0 & \WA21 \\
conflict for source                 & \WA48 & 0 &\WA6 &  0 &  0 &  0 &  0 &  0 &  0 &  0 &  0 & 0 & \WA 5 \\
conflict for bytecode               &     0 & 0 &   0 &  0 &  0 &  0 &  0 &  0 &  0 &  0 &  0 & 0 & \WA 2 \\
conflict for runtime code           & \WA55 & 0 &   0 &  0 &  0 &  0 &\WA3&  0 &  0 &  0 &  0 & 0 & \WA 2 \\
\midrule
\rowcolor{grey1} duplicate own id   &     0 & 0 &   0 &  0 &  2 &  0 &  0 &  0 &  0 &  0 &  0 & 0 &    70 \\
\rowcolor{grey1} duplicate address  &     0 & 0 &   0 &  0 &  0 &  0 &  0 &  0 &  0 &  0 &  0 & 0 &2\,299 \\
\rowcolor{grey1} duplicate source   &   807 & 0 &  24 &  3 &  3 & 44 &  0 &  0 & 30 &  0 &  7 & 1 &   572 \\
\rowcolor{grey1} duplicate bytecode &     0 & 3 &   5 &  0 &  0 &  8 &  0 &  0 &  0 & 10 &  0 & 0 &    34 \\
\rowcolor{grey1} duplicate runtime  &    46 & 1 &   5 &  3 &  0 &  0 &  0 &  0 &  1 &  6 &  0 & 0 &    34 \\
\midrule
\midrule
assessments ignored                 & 1\,330&  4&  40 & 18 & 16 & 52 &  3 &  0 & 31 & 16 &  7 &  1& 3\,210\\
assessments retained                &10\,410&375& 279 &702 &111 &292 &165 & 34 &219 &129 &343 &116& 7\,323\\
\bottomrule
\end{tabular}
\end{table}

\begin{table}
  \centering
  \caption{Criteria for Data Quality Assessment.}
  \label{tab:criteria}
  \setlength\tabcolsep{2pt}
  \begin{tabular}{llcp{8.5cm}}
    \toprule
    & Aspect & Score & Criterion \\
    \midrule
    \multirow{12}{*}{\SW{Accuracy}} 
    & \multirow{1}{*}{completeness} && table~\ref{tab:gt_contents} lists contents of the included GT sets \\
    && \good & source and bytecode are provided (and addresses if any) \\
    && \med &  some source and bytecodes provided \\
    && \bad & no code files are provided \\ 
    \\
    & \multirow{1}{*}{irredundancy} && table~\ref{tab:ignored} shows the duplicate entries in gray \\
    && \good & duplicates [\%] $\le 1$ \\
    && \med & $1 >$ duplicates [\%] $<10$ \\
    && \bad & duplicates [\%] $\ge 10$ (without bytecode) \\
    \\
    & \multirow{1}{*}{consistency} &&  \\
    && \bad & table~\ref{tab:ignored} shows contradictions and conflicts \\
    && \med & tables~\ref{tab:disagree} or \ref{tab:checked} show inconsistent or incorrect assessments \\
    && \good & otherwise \\
    \midrule
    \multirow{10}{*}{\SW{Relevance}}
    & heterogeneity && sections~\ref{ssec:variability} and \ref{ssec:quality} detail the heterogeneity of the GT sets \\
    && \bad &{All sets show low heterogeneity.} \\
    \\
    & data quantity && table~\ref{tab:gt_included} lists the number of assessments $A$ and weaknesses $W$ \\
    && \good & $A > 1000 \land W > 5$ \\
    && \bad  & $A <  500 \land W < 5$ \\
    && \med  & otherwise \\
    \\
    & timeliness && section ~\ref{ssec:variability} \\
    && \bad & {All sets lack recent contracts.}\\
    \bottomrule
  \end{tabular}
\end{table}

\clearpage
\section{Mapping to SWC Classes}\label{sec:mappings}
As the main taxonomy for weakness classification, we employ the one laid out in the SWC registry\footnote{\url{https://swcregistry.io}}.
In addition, we also provide the mapping to the taxonomy introduced in the DASP Top 10\footnote{\url{dasp.co}}.
For each included GT set, we map, whenever possible, its properties to appropriate classes in both taxonomies (table~\ref{tab:mapping}).
As SBcurated was classified with the rather coarse DASP taxonomy, we split the class ``DASP -- Access control'' into six SWC classes (105, 106, 112, 115, 118, 124)) and ``DASP -- DOS'' into two SWC classes (113, 128).

\begin{center}\small
\begin{longtable}{rrlp{7.5cm}}
\caption[Mapping of Wild GT Sets]{Mapping of Wild GT Sets.}
\label{tab:mapping} \\

\toprule
\multicolumn{1}{r}{\textbf{DASP}} & \multicolumn{1}{r}{\textbf{SWC}} & \multicolumn{1}{l}{\textbf{Set}} & \multicolumn{1}{l}{\textbf{Property}} \\ 
\midrule
\endfirsthead

\multicolumn{4}{c}%
{{\bfseries \tablename\ \thetable{} -- continued from previous page}} \\
\toprule
\multicolumn{1}{r}{\textbf{DASP}} & \multicolumn{1}{r}{\textbf{SWC}} & \multicolumn{1}{l}{\textbf{Sets}} & \multicolumn{1}{l}{\textbf{Property}} \\ 
\midrule 
\endhead

\midrule &&&\multicolumn{1}{r}{{Continued on next page}} \\
\endfoot

\bottomrule
\endlastfoot

 6 & 120 & CodeSmells & Block Info Dependency \\
10 &   0 & JiuZhou & byte{[}{]} \\
10 & 103 & CodeSmells & Compiler Version not fixed \\
10 &   0 & CodeSmells & Deprecated APIs \\
 5 & 113 & CodeSmells & Dos Under external influence \\
10 & 997 & CodeSmells & Greedy Contract \\
10 &   0 & CodeSmells & Hard Code Address \\
10 &   0 & CodeSmells & High Gas Consumption Data Type \\
10 &   0 & CodeSmells & High Gas Consumption Function Type \\
10 &   0 & CodeSmells & Misleading Data Location \\
10 &   0 & CodeSmells & Missing Interrupter \\
10 &   0 & CodeSmells & Missing Reminder \\
10 &   0 & CodeSmells & Missing Return statement \\
 5 & 128 & CodeSmells & Nest Call \\
 1 & 107 & CodeSmells & Reentrancy \\
10 & 132 & CodeSmells & strict balance equality \\
 2 & 115 & CodeSmells & Transaction state Dependency \\
 4 & 104 & CodeSmells & Unchecked External Call \\
10 &   0 & CodeSmells & Unmatched ERC-20 standard \\
10 &   0 & CodeSmells & Unmatched type assignment \\
10 & 135 & CodeSmells & Unused statement \\
 2 & 112 & ContractFuzzer & delegatecall\_dangerous \\
 4 & 104 & ContractFuzzer & exception\_disorder \\
10 & 997 & ContractFuzzer & freezing\_ether \\
 5 & 134 & ContractFuzzer & gasless\_send \\
 6 & 120 & ContractFuzzer & numberdependency \\
 1 & 107 & ContractFuzzer & reentrancy \\
 8 & 116 & ContractFuzzer & timedependency \\
 4 & 104 & Doublade & lowlevelcall\_result \\
 1 & 107 & Doublade & reentrancy \\
10 & 132 & Doublade & selfdestruct\_result \\
 2 & 115 & Doublade & tx\_result \\
 5 & 113 & Doublade & unexpectedrevert\_result \\
 1 & 107 & eThor & reentrancy \\
 7 & 114 & EthRacer & EO offchain \\
 7 & 114 & EthRacer & EO onchain \\
10 &   0 & EverEvolvingG. & airdrop-hunting \\
10 &   0 & EverEvolvingG. & call-after-destruct \\
10 & 996 & EverEvolvingG. & honeypot \\
 3 & 101 & EverEvolvingG. & integer-overflow \\
 1 & 107 & EverEvolvingG. & reentrancy \\
10 & 127 & JiuZhou & Any type of specified function variable \\
10 &   0 & JiuZhou & Change the contract status in the view or constant function \\
 5 &   0 & JiuZhou & Complex fallback function \\
10 &   0 & JiuZhou & continue in do-while \\
 5 & 128 & JiuZhou & DOS by gaslimit \\
 5 & 113 & JiuZhou & DOS by non-existent address or contract \\
10 & 132 & JiuZhou & Forced accept ethers \\
10 & 133 & JiuZhou & Hash Collisions With Multiple Variable Length Arguments \\
10 &   0 & JiuZhou & Hidden built-in symbols \\
10 & 119 & JiuZhou & Hide state variables \\
10 & 108 & JiuZhou & Implicit visibility level \\
10 & 110 & JiuZhou & Improper use of assert \\
10 & 123 & JiuZhou & Improper use of require \\
10 &   0 & JiuZhou & Improper use of revert \\
10 & 125 & JiuZhou & Incorrect inheritance order \\
 3 & 999 & JiuZhou & Integer division \\
 3 & 101 & JiuZhou & integer overflow and underflow \\
 3 & 999 & JiuZhou & Integer signedness \\
 3 & 999 & JiuZhou & Integer truncation \\
10 &   0 & JiuZhou & Invariant in loop \\
10 &   0 & JiuZhou & Invariant is not declared constant \\
10 & 997 & JiuZhou & Locked money \\
10 & 136 & JiuZhou & Non-public variables are accessed by public or external function \\
10 &   0 & JiuZhou & Non-standard naming \\
10 &   0 & JiuZhou & Nonstandard token interface \\
10 & 132 & JiuZhou & Pre-sent ether \\
10 & 136 & JiuZhou & Public data \\
 6 & 120 & JiuZhou & Randomness affected by miners \\
 1 & 107 & JiuZhou & Re-entrancy vulnerability \\
10 &   0 & JiuZhou & Removes dynamic array elements \\
 2 & 117 & JiuZhou & Replay attack \\
10 & 130 & JiuZhou & Right-To-Left-Overridecontrolcharacter (U+202E) \\
 9 & 995 & JiuZhou & Short address attack \\
10 &   0 & JiuZhou & Signature with wrong parameter \\
 2 & 124 & JiuZhou & Storage overlap attack \\
 2 & 106 & JiuZhou & Suicidal contract \\
 2 & 112 & JiuZhou & The call address or data are externally controlled \\
 8 & 116 & JiuZhou & Time affected by miners \\
10 &   0 & JiuZhou & too many digits \\
 7 & 114 & JiuZhou & Transaction order dependence \\
 2 & 115 & JiuZhou & Txorigin for authentication \\
 4 & 104 & JiuZhou & Unhandled exceptions \\
10 &   0 & JiuZhou & Uninitialized local variables \\
10 & 109 & JiuZhou & Uninitialized state variables \\
10 & 109 & JiuZhou & Uninitialized storage variable \\
10 & 103 & JiuZhou & Unlimited compiler versions \\
10 &   0 & JiuZhou & Unused public functions within a contract should be declared external \\
10 & 996 & JiuZhou & Use assembly code return in the constructor \\
10 & 111 & JiuZhou & Use deprecated built-in symbols \\
 2 & 105 & JiuZhou & Wasteful contract \\
 2 & 118 & JiuZhou & Write the wrong constructor name \\
10 & 129 & JiuZhou & Wrong operator \\
10 & 996 & NotSoSmartC & Honeypot Balance disorder \\
10 & 996 & NotSoSmartC & Honeypot Hidden state update \\
10 & 996 & NotSoSmartC & Honeypot Inheritance disorder \\
10 & 996 & NotSoSmartC & Honeypot Straw man contract \\
10 & 996 & NotSoSmartC & Honeypot Type overflow \\
10 & 996 & NotSoSmartC & Honeypot Uninitialised struct \\
10 &   0 & NotSoSmartC & Incorrect interface \\
 3 & 101 & NotSoSmartC & SWC-101 \\
 4 & 104 & NotSoSmartC & SWC-104 \\
 1 & 107 & NotSoSmartC & SWC-107 \\
 5 & 113 & NotSoSmartC & SWC-113 \\
 7 & 114 & NotSoSmartC & SWC-114 \\
 2 & 118 & NotSoSmartC & SWC-118 \\
10 & 119 & NotSoSmartC & SWC-119 \\
 6 & 120 & NotSoSmartC & SWC-120 \\
 2 & 124 & NotSoSmartC & SWC-124 \\
 5 & 128 & NotSoSmartC & SWC-128 \\
10 & 132 & NotSoSmartC & SWC-132 \\
 5 & 113 & NPChecker & Failed Call \\
 1 & 107 & NPChecker & Reentrancy \\
 6 & 120 & NPChecker & System Property Dependence \\
 7 & 114 & NPChecker & TOD \\
 4 & 104 & NPChecker & Unchecked Call \\
 2 & 105 & SBcurated & access\_control-105 \\
 2 & 106 & SBcurated & access\_control-106 \\
 2 & 112 & SBcurated & access\_control-112 \\
 2 & 115 & SBcurated & access\_control-115 \\
 2 & 118 & SBcurated & access\_control-118 \\
 2 & 124 & SBcurated & access\_control-124 \\
 3 & 101 & SBcurated & arithmetic \\
 6 & 120 & SBcurated & bad\_randomness \\
 5 & 113 & SBcurated & denial\_of\_service-113 \\
 5 & 128 & SBcurated & denial\_of\_service-128 \\
 7 & 114 & SBcurated & front\_running \\
10 & 109 & SBcurated & other-109 \\
 1 & 107 & SBcurated & reentrancy \\
 9 & 995 & SBcurated & short\_addresses \\
 8 & 116 & SBcurated & time\_manipulation \\
 4 & 104 & SBcurated & unchecked\_low\_level\_calls \\
 3 & 101 & SolidiFI & Overflow-Underflow \\
 1 & 107 & SolidiFI & Re-entrancy \\
 8 & 116 & SolidiFI & Timestamp-Dependency \\
 7 & 114 & SolidiFI & TOD \\
 2 & 115 & SolidiFI & tx.origin \\
 2 & 105 & SolidiFI & Unchecked-Send \\
 4 & 104 & SolidiFI & Unhandled-Exceptions \\
10 & 100 & SWCregistry & SWC-100 \\
 3 & 101 & SWCregistry & SWC-101 \\
10 & 102 & SWCregistry & SWC-102 \\
10 & 103 & SWCregistry & SWC-103 \\
 4 & 104 & SWCregistry & SWC-104 \\
 2 & 105 & SWCregistry & SWC-105 \\
 2 & 106 & SWCregistry & SWC-106 \\
 1 & 107 & SWCregistry & SWC-107 \\
10 & 108 & SWCregistry & SWC-108 \\
10 & 109 & SWCregistry & SWC-109 \\
10 & 110 & SWCregistry & SWC-110 \\
10 & 111 & SWCregistry & SWC-111 \\
 2 & 112 & SWCregistry & SWC-112 \\
 5 & 113 & SWCregistry & SWC-113 \\
 7 & 114 & SWCregistry & SWC-114 \\
 2 & 115 & SWCregistry & SWC-115 \\
 8 & 116 & SWCregistry & SWC-116 \\
 2 & 117 & SWCregistry & SWC-117 \\
 2 & 118 & SWCregistry & SWC-118 \\
10 & 119 & SWCregistry & SWC-119 \\
 6 & 120 & SWCregistry & SWC-120 \\
10 & 123 & SWCregistry & SWC-123 \\
 2 & 124 & SWCregistry & SWC-124 \\
10 & 125 & SWCregistry & SWC-125 \\
 5 & 126 & SWCregistry & SWC-126 \\
10 & 127 & SWCregistry & SWC-127 \\
 5 & 128 & SWCregistry & SWC-128 \\
10 & 129 & SWCregistry & SWC-129 \\
10 & 130 & SWCregistry & SWC-130 \\
10 & 131 & SWCregistry & SWC-131 \\
10 & 132 & SWCregistry & SWC-132 \\
10 & 133 & SWCregistry & SWC-133 \\
 5 & 134 & SWCregistry & SWC-134 \\
10 & 135 & SWCregistry & SWC-135 \\
10 & 136 & SWCregistry & SWC-136 \\
 6 & 120 & Zeus & Blk\_State\_Dep \\
 5 & 113 & Zeus & Failed\_send \\
 3 & 101 & Zeus & Int\_overflow \\
 1 & 107 & Zeus & Reentrancy \\
 7 & 114 & Zeus & Tx\_Order\_Dep \\
 2 & 115 & Zeus & Tx\_State\_Dep \\
 4 & 104 & Zeus & Unchkd\_send

\end{longtable}
\end{center}

\end{document}